\documentclass[article,aps,nofootinbib,showpacs,amsfonts,epsf]{revtex4}

\input{epsf.tex}

\newcommand{\E}{{\cal{E}}}
\newcommand{\s}{\sigma}

\renewcommand{\a}{\alpha}
\renewcommand{\k}{\kappa}

\newcommand{\be}{\begin{equation}}
\newcommand{\ee}{\end{equation}}
\newcommand{\bea}{\begin{eqnarray}}
\newcommand{\eea}{\end{eqnarray}}
\newcommand{\ba}{\begin{array}}
\newcommand{\ea}{\end{array}}
\def\J#1#2#3#4{{#1} {\bf #2}, #3 (#4)}
\def\PRD{Phys. Rev. D}
\def\PR{Phys. Rev.}
\def\PRL{Phys. Rev. Lett.}
\def\PTP{Prog. Theor. Phys.}

\def\APN{Ann. Phys. (NY)}

\def\AJ{Astrophys. J.}

\def\MNRAS{Mon. Not. R. Astron. Soc.}
\def\JMP{J. Math. Phys.}
\def\CPAM{Comm. Pure Appl. Math.}

\def\CQG{Class. Quantum Grav.}

\def\PLA{Phys. Lett. A}
\def\PLB{Phys. Lett. B}

\begin{document}
\draft
\title{Binary systems of recoiling extreme Kerr black holes}

\author{V.~S.~Manko,$^\dagger$ E.~Ruiz$^\ddagger$ and M. B. Sadovnikova$^\sharp$}
\address{$^\dagger$Departamento de F\'\i sica, Centro de Investigaci\'on y
de Estudios Avanzados del IPN, A.P. 14-740, 07000 Ciudad de
M\'exico, Mexico\\$^\ddagger$Instituto Universitario de F\'{i}sica
Fundamental y Matem\'aticas, Universidad de Salamanca, 37008
Salamanca, Spain\\$^\sharp$Department of Quantum Statistics and
Field Theory, Lomonosov Moscow State University, Moscow 119899,
Russia}

\begin{abstract}
In the present paper the repulsion of two extreme Kerr black holes
arising from their spin-spin interaction is analyzed within the
framework of special subfamilies of the well-known
Kinnersley-Chitre solution. The binary configurations of both
equal and nonequal extreme repelling black holes are considered.
\end{abstract}


\maketitle

\section{Introduction}

In a recent paper \cite{MRu} the possibility of the repulsion of
two equal subextreme Kerr black holes due to their spin-spin
interaction has been discovered within the framework of the
extended solitonic spacetimes \cite{MRu2,MRS}. Since one may
reasonably suppose that the repulsion effect could be not less,
but probably even stronger, in the case of extreme constituents,
it would certainly be of interest to supplement the research
reported in \cite{MRu} with the study of the binary configurations
composed of extreme Kerr black holes. Such configurations
naturally arise, taking into account that the extreme limit of the
usual double-Kerr metric \cite{KNe} is given by the well-known
Kinnersley-Chitre solution \cite{KCh}, as special subfamilies of
the latter solution describing two extreme black holes separated
by a massless strut, and these were first identified and discussed
in the paper \cite{MRu3}. In a later work \cite{MRSa} the binary
systems of extreme black holes were analyzed in more detail, and
it might be noted that we overlooked the repulsion effect in the
second subfamily from \cite{MRSa}, being convinced that the
interaction force in that subfamily had to be a positive quantity.
At the same time, we did not pay much attention to the repulsion
of black holes in one very special subcase of the
Kinnersley-Chitre solution analyzed in \cite{MRSa} since its
binary configurations had the ratio of total angular momentum to
total mass analogous to the one characterizing a single
hyperextreme Kerr source~\cite{Ker}, i.e. $|J|/M^2>1$, which
looked to us very exotic.

The objective of the present paper, in light of the aforementioned
discovery of the black hole repulsion, is to single out and
discuss the previously overlooked binary configurations of both
equal and nonequal {\it repelling extreme} Kerr black holes that
arise from the Kinnersley-Chitre solution. In particular, we are
going to give a novel representation of the metric function
$\omega$ for the binary systems with struts which is simpler than
the one considered in \cite{MRu3,MRSa} and makes evident the
presence of the axis between the constituents; we will also derive
concise analytic expressions for the interaction force and for the
norm of the axial Killing vector. The application of the latter
expression to the analysis of the geometry around the extreme
sources will allow us to obtain important information about the
characteristic features of the physically meaningful
configurations of repelling black holes endowed with positive
masses.

\section{Repulsion of two identical extreme Kerr black holes}

The first subfamily of the Kinnersley-Chitre solution that we are
going to consider describes {\it identical} extreme corotating
Kerr black holes separated by a massless strut. This subfamily was
identified in \cite{MRu3} and it represents the extreme limit of
the solution \cite{MRu}. Its Ernst complex potential \cite{Ern} is
determined by the expression \cite{MRu3}
\bea \E&=&(A-B)/(A+B), \nonumber\\
A&=&p^2(x^4-1)+q^2(y^4-1)-\beta^2(x^2-y^2)^2
-2ixy[pq(x^2-y^2)+\beta(x^2+y^2-2)], \nonumber\\
B&=&2[\beta(qx+ipy)(x^2-y^2)+px(x^2-1)+iqy(y^2-1)],
\label{eps_sim} \eea
where
\be \beta=\frac{1}{2p}[\Delta_{\rm S}+q(1+p)], \quad \Delta_{\rm
S}=\sqrt{(1+p)(1+3p^2+pq^2)}, \label{b_sim} \ee
and the real parameters $p$ and $q$ are subject to the constraint
$p^2+q^2=1$. The prolate spheroidal coordinates $x$ and $y$ are
related to the Weyl-Papapetrou cylindrical coordinates $\rho$ and
$z$ by the formulas
\be x=\frac{1}{2\k}(r_++r_-), \quad y=\frac{1}{2\k}(r_+-r_-),
\quad r_\pm=\sqrt{\rho^2+(z\pm\k)^2}, \label{xy} \ee
$\kappa$ being a positive real constant; the inverse
transformation is
\be \rho=\kappa\sqrt{(x^2-1)(1-y^2)}, \quad z=\kappa xy.
\label{rz} \ee
On the symmetry axis defined by the points $\{\rho=0$,
$-\infty<z<+\infty\}$, potential (\ref{eps_sim}) takes the form
(for $z>\kappa$)
\bea &&\E(\rho=0,z)={e_-}/{e_+}, \nonumber\\
&&e_\pm=(p^2-\beta^2)z^2+2\kappa[\pm(p+q\beta)-i(pq+\beta)]z
+\kappa^2(p^2+\beta^2\pm 2ip\beta). \label{ad} \eea

The corresponding metric functions $f$, $\gamma$ and $\omega$
entering the line element
\bea d s^2&=&\kappa^2f^{-1}\left[e^{2\gamma}(x^2-y^2)
\left(\frac{d x^2}{x^2-1}+\frac{d
y^2}{1-y^2}\right)+(x^2-1)(1-y^2)d\varphi^2\right]-f(d t-\omega
d\varphi)^2, \nonumber\\ \label{Pap} \eea
are given by the expressions
\bea f&=&\frac{E}{D}, \quad e^{2\gamma}
=\frac{E}{K_0^2(x^2-y^2)^4},
\quad \omega=\frac{\k(x-1)(y^2-1)F}{E}, \nonumber\\
E&=&\mu^2+(x^2-1)(y^2-1)\s^2, \nonumber \\
D&=&E+\mu\nu-(x-1)(y^2-1)\s\tau,
\nonumber\\ F&=&(x+1)\s\nu+\mu\tau, \nonumber\\
\mu&=&p^2(x^2-1)^2+q^2(y^2-1)^2-\beta^2(x^2-y^2)^2,
\nonumber\\
\s&=&2[pq(x^2-y^2)+\beta(x^2+y^2)], \nonumber\\ \nu&=&(4/K_0)\{K_0
[px(x^2+1)+2x^2] +K_0 q\beta x(x^2-y^2)
-2p^2\beta^2(x^2-y^2) \nonumber\\
&&+4\beta(pq+\beta)x^2\}, \nonumber\\
\tau&=&\frac{4}{p^2-q^2}\{(p^2-q^2)[q(x+y^2)
+p\beta(x^2-y^2)] -(p\beta-q-2pq)(x+1)\}, \nonumber\\
K_0&=&p^2-\beta^2, \label{mf_sim} \eea
where the novel form of $\omega$ contains the factor $(x-1)$
explicitly, which means that $\omega$ vanishes when $x=1$, i.e. on
the intermediate ($|z|<\kappa$) part of the symmetry axis (see
Fig.~1). This in turn implies that the latter portion of the axis
is a strut \cite{BWe,Isr} -- a conical angle deficit (or excess)
defined by the axis value of the metric function $\gamma$.

The Komar \cite{Kom} mass $M$ and angular momentum $J$ of each
extreme Kerr constituent are given by the formulas
\be M=\frac{\kappa (p+q\beta)}{p^2-\beta^2}, \quad
J=\frac{M[2(pq+\beta)M+\kappa p\beta]}{p+q\beta}, \label{MJ_sim}
\ee
so that the angular momentum-mass ratio has the form
\be \delta_S\equiv\frac{J}{M^2}=\frac{1}{2} [(2 - p)\Delta_S-
pq(3-p)], \label{delS} \ee
while the total values $M_T$ and $J_T$ are just twice the
respective individual quantities: $M_T=2M$ and $J_T=2J$. In the
paper \cite{MRu3} it was established that there are two parameter
ranges at which $M$ takes positive values, namely,
\be -\frac{1}{\sqrt{2}}<p<0, \quad q>0 \quad \mbox{and} \quad
0<p<\frac{1}{\sqrt{2}}, \quad q<0, \label{pq_sim} \ee
and these sets of parameters define, as will be seen below, two
physically distinct binary configurations of extreme identical
Kerr black holes, the first one (with negative $p$) describing a
pair of black holes {\it attracting} each other, and the second
configuration (with positive $p$) describing a pair of {\it
repelling} black holes. Indeed, the interaction force between two
constituents is determined by the formula \cite{Isr,Wei}
\be {\cal F}= \frac{1}{4}\left(e^{-\gamma_0}-1\right), \label{F}
\ee
where $\gamma_0$ denotes the value of the metric function $\gamma$
on the part of the axis separating the black holes. Then for both
of the aforementioned subfamilies, formulas (\ref{mf_sim}) yield
the same simple expression
\be {\cal F}= \frac{p^2-q^2}{4(q^2-\beta^2)}, \label{F_sim} \ee
and the plots in Fig.~2 show that ${\cal F}$ takes positive values
in the case of the first subfamily, and negative values in the
case of the second subfamily of binary configurations. Therefore,
in the systems with $p<0$, gravitational attraction overcomes
spin-spin repulsion, while in the systems with $p>0$, spin-spin
repulsion overcomes gravitational attraction, which confirms our
above interpretations given to the two subfamilies.

An important point to emphasize here is that $1<\delta_S<2$
independently of whether a binary configuration represents
attracting or repelling black holes, as it follows from the plots
in Fig.~3. This means, on the one hand, that the ratio $\delta_S$
of all extreme black holes in both subfamilies exceeds the
analogous value (equal to 1) of a single extreme Kerr black hole,
and, on the other hand, that there are configurations of
attracting and repelling extreme black holes sharing any
prescribed particular value of $\delta_S$. The latter might look
strange at first glance but actually has a simple explanation --
the configurations from different subfamilies possessing the same
$\delta_S$ and the same coordinate separation distance $2\kappa$
will have different masses, the repelling black holes carrying the
larger mass, which would be physically equivalent to having two
binary systems with the same $\delta_S$ and masses but different
separation distances, the shorter distance obviously corresponding
to the repelling black holes. For example, it is easy to see that
if $\delta_S=1.4$ and $\kappa=2$, then $p\simeq0.198$,
$q\simeq-0.98$, and the repelling black holes will have the mass
$M\simeq13.11$ and angular momentum $J\simeq240.628$. On the other
hand, from (\ref{MJ_sim}) and (\ref{delS}) it follows that the
binary system of attracting black holes with the latter values of
mass and angular momentum is characterized by $p\simeq-0.282$,
$q\simeq0.96$ and a considerably larger separation parameter
$\kappa\simeq33.075$. This result is logic if one recalls that the
spin-spin repulsion force is inversely proportional to $\kappa^4$
\cite{Wal} and hence decreases more rapidly with a larger distance
than the gravitational force.

In the paper \cite{MRu3} it was observed that the stationary limit
surface (SLS) of a solution with positive $p$ has some features
that distinguish it from the SLS of a solution with negative $p$.
The main feature is of course the presence of a massless ring
singularity outside the symmetry axis in the former solution, the
appearance of which was attributed in \cite{MRu3} to the
instabilities during the merging of SLSs. However, such an
interpretation does not look quite precise in view of the
intrinsic nature of the solutions with positive $p$ established in
the present paper -- repelling black holes -- and so we find it
instructive in what follows to reexamine in more detail the
geometrical properties of the solutions from the two subfamilies.
For completeness, it is also desirable to study the issue of
possible appearance of the regions with closed timelike curves
(CTCs) attached to the massless ring singularities in order to
evidence their benign character. To fulfil the latter objective we
have obtained a very simple representation for the norm
$\eta^\a\eta_\a$ of the axial Killing vector, namely,
\bea \eta^\a\eta_\a&=&
\k^2(x-1)(1-y^2){N}/{D}, \nonumber\\
N&=&(x+1)(\mu+\nu)^2-(x-1)(1-y^2)(x\s+\s-\tau)^2, \label{norm}
\eea
whose negative values determine the regions with CTCs.

In Fig.~4 we have plotted the SLSs for the particular two
solutions with $\delta_S=1.4$ from the above example. The SLS in
Fig.~4(a) belongs to the configuration of attracting extreme black
holes separated by the coordinate distance $2\kappa\simeq66.149$,
and it does not have any massless ring singularity off the
symmetry axis or a region of CTCs. In contrast, the SLS in
Fig.~4(b) is accompanied by a massless ring singularity located in
the equatorial plane, and also by a region of CTCs (inside the
dotted curve) touching that singularity. It is clear that the
extreme constituents in the second binary configuration are
situated very close to each other, their SLSs having already
formed a common SLS. However, since we have established that the
black holes in Fig.~4(b) are repelling, it would be now plausible
to infer that the ring singularity is not a product of merging of
two SLSs, but rather a result of the beginning of desintegration
of the common SLS into two parts.

Though the above analysis of Fig.~4 may look to provide a clear
and simple description of how the attraction and repulsion of the
extreme black holes work, the real situation with the interaction
force is far more interesting and even puzzling. To see this, let
us consider another two configurations of attracting and repelling
black holes with the same separation parameter $\kappa=2$ and
angular-momentum--mass ratio $\delta_S=1.99$, for which we readily
find from (\ref{MJ_sim}) and (\ref{delS}) that the attracting
constituents have the mass $M\simeq13.978$ and angular momentum
$J\simeq388.792$, while the mass and angular momentum of the
repelling constituents are, respectively, $M\simeq30.714$ and
$J\simeq1877.263$. In Fig.~5 we have plotted the SLSs of these
binary systems, and it can be seen from Fig.~5(a) that the
attracting extreme black holes have formed a common SLS, and
neither a massless ring singularity nor a region of CTCs appear on
that figure, which means that merging of the individual SLSs in
that configuration has been realized through a smooth analytic
process. At the same time, the SLS of repelling black holes
depicted in Fig.~5(b), like earlier the SLS in Fig.~4(b), is
accompanied by a massless ring singularity and by a region with
CTCs, though it might look strange that the latter region is
smaller than in Fig.~4(b) despite the larger angular momentum of
the new system compared to the configuration in Fig.~4(b).
However, a more exciting question would be about the binary
systems in Figs.~4(b) and 5(a): the separation of extreme
constituents in both systems is the same, so why do the black
holes in Fig.~5(a) attract each other (instead of repelling) if
their mass and the ratio $\delta_S$ are even larger than in the
configuration from Fig.~4(b) and consequently are expected to
produce a greater spin-spin repulsion effect?

Though a possible explanation for a smaller CTC region in
Fig.~5(b) could be that the SLS of that configuration is only at
the beginning of its splitting into two parts, while in Fig.~4(b)
the division of the SLS is at a more advanced stage, the answer to
the second question is not that simple and actually seems to be
related to the recent findings of the paper \cite{MRu}. First of
all, Fig.~5(a) demonstrates that the unification of two SLSs can
be a smooth process even at small separation distances and large
values of mass, angular momentum and parameter $\delta_S$
approaching 2, so that the repulsion effect taking place in the
configuration depicted in Fig.~4(b) should be basically attributed
to instabilities of the SLS. However, the latter instabilities
cannot be entirely explained by the SLS splitting due to repulsion
of black holes alone, simply because an analogous repulsion does
not take place inside the configuration from Fig.~5(a). Therefore,
we inevitably arrive at the conclusion that the binary system from
Fig.~4(b) represents a sort of a resonant state that produces
instabilities of the SLS, as well as the configuration from
Fig.~5(b). Such an inference looks plausible since a configuration
of attracting black holes and a configuration of repelling black
holes cannot have simultaneously the same particular values of
$M$, $J$ and $\kappa$, so that a resonant state producing the SLS
instability occurs only when the latter characteristics of a
binary system of extreme constituents all achieve the particular
values belonging to the subfamily of repelling extreme black
holes, i.e. a configuration with some admissible $M$ and $J$ must
have a concrete parameter $\kappa$, a configuration with known $M$
and $\kappa$ -- the concrete angular momentum $J$, and a
configuration with some given $J$ and $\kappa$ -- a special value
of $M$ for producing a resonant state and repulsion.

One of the sources of the SLS instability could be the
nonuniqueness of the binary black-hole configurations
characterized by the same mass and angular momentum of the
constituents discovered in \cite{MRu}, because the SLS in this
case must be affected by a possible spontaneous change of the
multipole structure in such configurations. This sort of
instability is certainly proper of the configuration from
Fig.~5(b), which can be shown to belong to the so-called
``triangle zone'' of nonuniqueness \cite{MRu,MRu4} giving rise to
three different configurations of black holes: apart from the
configuration of extreme black holes from Fig.~5(b), there are two
other configurations of nonextreme identical black holes with the
same mass $M\simeq30.714$ and angular momentum $J\simeq1877.263$
of the constituents, whose respective rescaled dimensionless
quantities $\s$ defining half-lengths of the horizons have been
found to be $\s\simeq0.215$ and $\s\simeq0.175$.

At the same time, it is not difficult to check that the
nonuniqueness argument is not applicable to the configuration of
repelling extreme black holes from Fig.~4(b), as the latter does
not belong to the nonuniqueness zone and hence there are no other
binary configurations with the same mass, angular momentum and
separation distance. The instability, notwithstanding, could be
simply a result of a special, resonant status of the configuration
itself which might produce perturbations of the SLS, and these in
turn could give rise to the repulsion effect.

\section{Repulsion of two unequal extreme Kerr black holes}

We now turn to consideration of the binary configurations of
unequal extreme Kerr black holes generalizing the equatorially
symmetric binary systems from the previous section. The Ernst
complex potential $\E$ defining this subfamily of the
Kinnersley-Chitre solution has the form \cite{MRu3,MRSa}
\bea \E&=&(A-B)/(A+B), \nonumber\\
A&=&p^2P^2\{p^2(x^4-1)+q^2(y^4-1)
-2ixy[pq(x^2-y^2)+\beta(x^2+y^2-2)]\} \nonumber\\
&&+[Q^2(p+q\beta)^2-p^2P^2\beta^2](x^2-y^2)^2
+2ipPQ(p+q\beta)(x^2+y^2-2x^2y^2), \nonumber\\
B&=&2pP(P-iQ)\{(x^2-y^2)[pP\beta(qx+ipy)+Q(p+q\beta)(qy+ipx)]
\nonumber\\ &&+pP[px(x^2-1)+iqy(y^2-1)]\}, \label{eps_gen} \eea
where
\be \beta=\frac{p[P\Delta+q(1+pP+Q^2)]}{2(p^2-Q^2)}, \quad
\Delta=\sqrt{4p^2(1+pP)+q^2(p+P)^2}, \label{b_2} \ee
and the real parameters $P$ and $Q$ are subject to the same
constraint as $p$ and $q$: $P^2+Q^2=1$.

The corresponding metric functions $f$, $\gamma$ and $\omega$,
with a new form of $\omega$, are given by the expressions
\bea f&=&\frac{E}{D}, \quad e^{2\gamma}
=\frac{E}{K_0^2(x^2-y^2)^4},
\quad \omega=\frac{\k(x-1)(y^2-1)F}{E}, \nonumber\\
E&=&\mu^2+(x^2-1)(y^2-1)\s^2, \nonumber \\
D&=&E+\mu\nu-(x-1)(y^2-1)\s\tau,
\nonumber\\ F&=&(x+1)\s\nu+\mu\tau, \nonumber\\
\mu&=&p^2P^2[p^2(x^2-1)^2+q^2(y^2-1)^2]
+[Q^2(p+q\beta)^2-p^2P^2\beta^2](x^2-y^2)^2,
\nonumber\\
\s&=&2pP\{pP[pq(x^2-y^2)+\beta(x^2+y^2)]+2Q(p+q\beta)xy\},
\nonumber\\ \nu&=&(4pP/K_0)\{K_0 pP[pPx(x^2+1)+2x^2+qQy(y^2+1)]
+K_0(x^2-y^2) \nonumber\\ &&\times
[(p^2Q^2+pq\beta)x+PQ(pq+\beta)y]
-2pP[q^2Q^2(p+q\beta)^2+p^4P^2\beta^2](x^2-y^2) \nonumber\\
&&+4p^2P^2(pq+\beta)x[pP\beta x+Q(p+q\beta)y)]\}, \nonumber\\
\tau&=&\frac{4pP}{p^2-q^2}\{pP(p^2-q^2)(x+1)(qPx-pQy)
+(p^2-q^2)[(p^2-Q^2)\beta-pq](x^2-y^2) \nonumber\\
&&-[(p^2-Q^2)\beta-pq(1+2pP)](x+1)\}, \nonumber\\
K_0&=&p^2P^2(p^2-\beta^2)+Q^2(p+q\beta)^2, \label{mf_gen} \eea
and one can see that $\omega$ automatically vanishes at $y=\pm1$
and $x=1$ (see Fig.~1). The formulas for the total mass and total
angular momentum of these binary configurations are
\bea M_T&=&\frac{2\kappa
p^2P(p+q\beta)}{p^2P^2(p^2-\beta^2)+Q^2(p+q\beta)^2}, \nonumber\\
J_T&=&M\left[\frac{P(pq+\beta)M}{p+q\beta}-\frac{\kappa}{p}
\left(qQ^2-\frac{p^2P^2\beta} {p+q\beta}\right)\right],
\label{MJ_gen} \eea
while for the individual Komar masses and angular momenta of the
constituents we have the expressions \cite{MRSa}\footnote{We have
rectified two misprints in the formulas (18) of \cite{MRSa}: the
denominators of $J_1$ and $J_2$ should not have the factor $p$.}
\bea M_1&=&\frac{\kappa[(q+pqP-p^2Q)\Delta-(1+pP)
(p+p^3+q^2P-pqQ)+pq^3Q]} {2p(1+pP)(p^2-q^2)}, \nonumber\\
M_2&=&\frac{\kappa[(q+pqP+p^2Q)\Delta
-(1+pP)(p+p^3+q^2P+pqQ)-pq^3Q]} {2p(1+pP)(p^2-q^2)}, \nonumber\\
J_1&=&\frac{(1+pP+qQ)M_1^2} {2(p+P)^2}
[(1+pP+q^2)\Delta-4pq+pq(p-P)^2], \nonumber\\
J_2&=&\frac{(1+pP-qQ)M_2^2} {2(p+P)^2}
[(1+pP+q^2)\Delta-4pq+pq(p-P)^2], \label{MiJi_2} \eea
and these were obtained as limits of the general expressions found
by Tomimatsu \cite{Tom} and Dietz and Hoenselaers \cite{DHo} for
the non-extended double-Kerr solution \cite{KNe}. Note that $J_1$
and $J_2$ are given in the form most suitable for evaluating the
individual angular-momentum--mass ratios $\delta_i=J_i/M_i^2$,
$i=1,2$. The subindex 1 refers to the upper black hole and the
subindex 2 to the lower black hole. As was observed in
\cite{MRSa}, $J_1$ and $J_2$ cannot have opposite signs, so the
extreme constituents are corotating. Reduction to the case of two
equal black holes occurs when $P=1$, $Q=0$.

The masses $M_1$ and $M_2$ take positive values, which are only of
interest to us, for the following ranges of the parameters
\cite{MRSa}:
\be -\frac{1}{\sqrt{2}}<p<0, \quad q>0; \quad
-1<P<-p\,\,\cup\,\,q<P<1, \label{pos_1} \ee
and
\be 0<p<\frac{1}{\sqrt{2}}, \quad q<0; \quad
-1<P<q\,\,\cup\,\,-p<P<1, \label{pos_2} \ee
for arbitrary sign of $Q$. Comparing (\ref{pos_1}) with the
negative values of $p$ in (\ref{pq_sim}), on the one hand, and
(\ref{pos_2}) with the positive values of $p$ in (\ref{pq_sim}),
on the other hand, one would anticipate that the subfamily of
unequal black holes defined by (\ref{pos_1}) must describe
attracting constituents, while the subfamily defined by
(\ref{pos_2}) must describe the repelling constituents. The real
situation is however a bit more complicated, as we shall see later
on.

The expression for the interaction force between two unequal
extreme Kerr constituents is obtainable from (\ref{F}) and has the
form
\be {\cal F}= \frac{p^2P^2(p^2-q^2)}
{4[p^2P^2(q^2-\beta^2)+Q^2(p+q\beta)^2]}, \label{F_as} \ee
and one can see that the sign of $Q$ in the above expression for
${\cal F}$ is irrelevant, unlike the sign of $q$. Note that the
norm of the axial Killing vector in the case of unequal
constituents is defined by the same simple formula (\ref{norm})
obtained in the previous section for identical black holes, but
the quantities $\mu$, $\nu$, $\s$ and $\tau$ entering it must be
taken this time from (\ref{mf_gen}).

The analysis of the interaction force (\ref{F_as}) reveals that
actually any of the subfamilies (\ref{pos_1}) or (\ref{pos_2}) may
describe configurations of both attracting and repelling unequal
extreme black holes. The case of attracting constituents in the
subfamily (\ref{pos_1}) is defined by the positive values of $P$
on the interval $P\in(q,1)$, while in the subfamily (\ref{pos_2})
such configurations correspond to negative $P$ on the interval
$P\in(-1,q)$, and these are of no interest to us. The
configurations of repelling constituents arise in the subfamily
(\ref{pos_1}) at $P\in(-1,-p)$, and in the subfamily (\ref{pos_2})
at $P\in(-p,1)$, the latter two intervals containing both positive
and negative values of $P$. Apparently, the case of unequal
constituents provides more possibilities for the extreme black
holes to repel each other, and the individual
angular-momentum--mass ratios $\delta_i$ of the repelling unequal
constituents may exceed significantly the respective maximum value
2 of identical black holes. Restricting ourselves to the repulsion
of unequal black holes, in Figs.~6 and 7 we have depicted the
SLSs, the regions with CTCs and massless ring singularities for
the binary configurations defined by the same values of $\kappa$,
$p$ and $q$ as in Figs.~4 and 5 but different values of $P$ and
$Q$ (we recall that for the identical black holes $P=1$, $Q=0$).
Thus, in Fig.~6(a) the values of $\kappa$, $p$ and $q$ defining
the system of repelling black holes are the same as in Fig.~4(b),
but with $P=0.5$, $Q\simeq0.866$. The masses of black holes are
$M_1\simeq9.918$ and $M_2\simeq6.194$, and the corresponding
individual angular-momentum--mass ratios are $\delta_1\simeq0.618$
and $\delta_2\simeq4.812$ (we give them instead of the angular
momenta of the constituents); though the splitting of the common
SLS into two disconnected parts has already occurred, a byproduct
of that non-smooth process is the appearance of the third separate
component of SLS that characterizes the instability zone marked by
the massless ring singularity and associated region of CTCs. The
interaction force ${\cal F}\simeq-0.2493$, hence being repulsive,
and the binary system as a whole looks to a distant observer as a
subextreme object because $\delta_T\equiv J_T/M_T^2\simeq0.945$.

The configuration from the subfamily (\ref{pos_1}) whose SLS is
plotted in Fig.~6(b) is defined by the same particular values of
$\kappa$, $p$ and $q$ as the configuration of attracting
constituents in Fig.~4(a), but its $P$ and $Q$ are different:
$P=-0.8$, $Q=0.6$, which leads to the repulsion of the
constituents since ${\cal F}\simeq-0.2487$. The above change of
$P$ and $Q$ affects drastically the masses of black holes,
yielding $M_1\simeq130.589$ and $M_2\simeq176.792$, together with
the angular-momentum--mass ratios $\delta_1\simeq2.776$ and
$\delta_2\simeq1.001$; at the same time, $\delta_T\simeq0.832$
which is characteristic of a subextreme object, as in the previous
example. The presence of a massless ring singularity and
associated region of CTCs is an indication that there is a zone of
instability due to disintegration of the common SLS.

The plots given in the next Figs.~7(a) and 7(b) are very similar,
despite representing configurations from different subfamilies
(\ref{pos_1}) and (\ref{pos_2}). Both configurations are composed
of repelling black holes, have large SLSs and small instability
zones accompanied by massless ring singularities and regions of
CTCs. The configuration from Fig.~7(a) shares the same values of
$\kappa$, $p$ and $q$ with the configuration from Fig.~5(b), but
its $P$ and $Q$ are different: $P=0.2$, $Q\simeq0.98$. The
individual masses $M_i$ and ratios $\delta_i$ of the constituents
are $M_1\simeq34.699$, $M_2\simeq9.04$, $\delta_1\simeq1.195$ and
$\delta_2\simeq5.811$, so that the ratio $\delta_T$ involving
total mass and total angular momentum becomes
$\delta_T\simeq1.00042$, and the binary system looks to a distant
observer as a hyperextreme object. On the other hand, the values
of $\kappa$, $p$ and $q$ of the configuration from Fig.~7(b) are
the same as for the configuration from Fig.~5(a), the remaining
two parameters having the values $P=-0.8$ and $Q=0.6$; the masses
of the constituents are $M_1\simeq29.431$ and $M_2\simeq50.172$,
and the corresponding angular-momentum--mass ratios are
$\delta_1\simeq2.814$ and $\delta_2\simeq1.545$, the ratio
$\delta_T$ of this configuration being equal to 0.998. These
examples clearly demonstrate that the configurations of unequal
extreme black holes lend more opportunities for their constituents
to repel each other than the binary systems comprised exclusively
of identical black holes.

\section{Discussion and conclusions}

Therefore, we have shown that there are vast families of binary
configurations of repelling extreme Kerr black holes endowed with
positive masses. The repulsion effect arises due to spin-spin
interaction of the constituents when a binary system achieves a
specific state, which we tentatively call ``resonant'',
characterized by instabilities of the SLS. Such instabilities seem
to be mainly determined by the disintegration of the common SLS
into two or more fragments through a non-smooth process, but could
also be an intrinsic characteristic and inseparable part of the
resonant states themselves, somehow contributing to the repulsion
of black holes too. In our analysis we have put emphasis on the
geometrical characteristics of the repelling extreme black holes
such as SLSs, massless ring singularities and regions with CTCs in
order to elucidate an important role of the latter two in
maintaining the stationarity of the configurations and to justify
their benign character. Since historically the massless ring
singularities puzzled the researchers for quite a long time, in
what follows we would like to make a few additional comments on
them.

While the physical significance of massless struts (conical
singularities) was understood long ago \cite{BWe,Isr} and the
presence of struts in the two-body solutions of general relativity
is considered as ``a very satisfactory feature of this nonlinear
theory'' \cite{SKM}, the massless ring singularities were first
reported by far later in relation to the Tomimatsu-Sato solutions
\cite{TSa} and were commonly regarded, together with the
associated regions of CTCs, as some undesirable pathological
characteristics of spacetimes \cite{GRu}. It was conjectured by
Hoenselaers \cite{Hoe} that in the multi-black-hole configurations
a constituent with negative mass should be accompanied by a naked
singularity, and in the paper \cite{Man} it was demonstrated that
the massless ring singularity in $\delta=2$ Tomimatsu-Sato
solution is linked to a region of negative mass existing in that
solution. Examples of appearance of ring singularities in the
spacetimes involving negative mass were also given in the
framework of extended double-Kerr solution \cite{MRS}, so that one
might think in principle that a naked singularity is a proper
intrinsic feature of {\it any} negative mass and then call
``unphysical'' both the massless ring singularity and the negative
mass (what usually happens).

However, the discovery of the binary black-hole configurations
involving exclusively positive masses and yet having ring
singularities outside the symmetry axis, such as the ones
considered in \cite{MRu,MRu3,MRSa}, urges us to make a more
profound and broader look at the massless ring singularities in
order to bring to light the universal task they fulfil in the
stationary axisymmetric spacetimes. Taking into account that the
Schwarzschild and Kerr solutions endowed with negative mass are
known to be unstable \cite{GDo,CCa}, and also that a massless ring
singularity arising in the configurations of interacting black
holes carrying positive masses can be associated with the
instabilities of SLSs, it would be plausible to suppose that the
unique function such singularities perform in stationary
spacetimes is simply maintaining the stationarity of the latter,
which is actually the same role as performed by the conical
singularities, but covering more situations where the
instabilities may occur. In this respect, it is clear that the
ring singularities and associated regions of CTCs in the
configurations of repelling black holes considered in the previous
sections are needed to detain the dynamical evolution of SLSs
which would otherwise go on changing their shapes despite the
unchanging positions of black holes on the symmetry axis ensured
by struts. It should be emphasized that in the real axisymmetric
nonstationary configurations of interacting extreme black holes
which are mimicked by our stationary solutions, neither the struts
on the axis nor the ring singularities off the axis of symmetry
will be present because the evolution of the black holes and SLSs
then will not be artificially constrained by the stationarity
condition. As a result, the repelling constituents will be moving
away from each other and the SLSs will be changing their aspect in
a natural, unrestricted way, the dynamical evolution being also
accompanied by a small loss of energy in the form of gravitational
waves \cite{DOr}.

Lastly, it is tempting to speculate that the observational
phenomenon of recoiling black holes \cite{KZL,BSK} which the
astronomers attribute so far to the powerful gravitational
radiation liberated during the merging process, could be just a
natural outcome of the spin-spin interaction of corotating black
holes during the head-on collision.

\section*{Acknowledgments}
This work was supported in part by the CONACYT of Mexico, and by
Project FIS2015-65140-P (MINECO/FEDER) of Spain. One of us (VSM)
would like to thank the Department of Quantum Statistics and Field
Theory of the Moscow State University for the hospitality extended
to him during his visit there in July 2018.

\newpage

\begin{figure}[htb]
\centerline{\epsfysize=80mm\epsffile{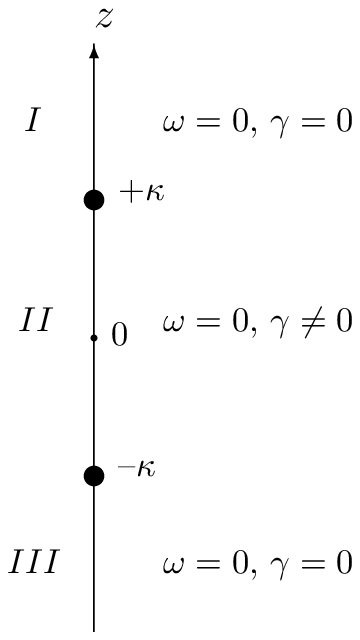}} \caption{Location
of two extreme black holes on the symmetry $z$-axis (at the points
$z=\pm\kappa$), and behavior of the metric functions $\omega$ and
$\gamma$ on three different parts of the symmetry axis: $z>\kappa$
(part $I$), $-\kappa<z<\kappa$ (part $II$) and $z<-\kappa$ (part
$III$).}
\end{figure}

\begin{figure}[htb]
\centerline{\epsfysize=50mm\epsffile{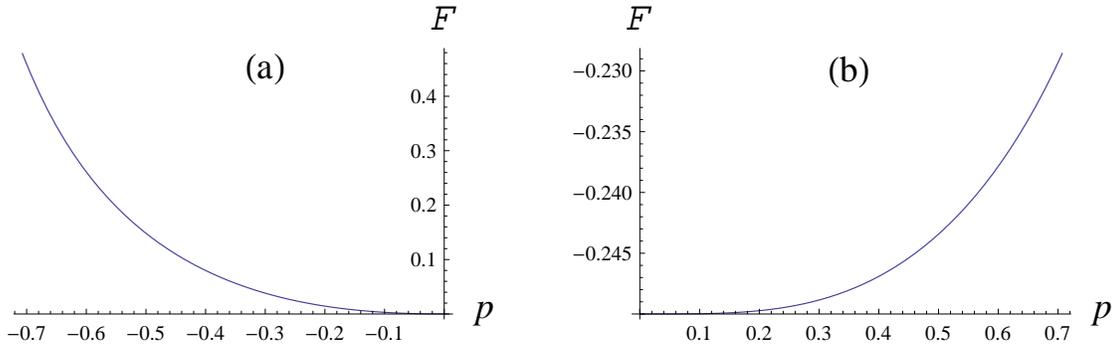}}
\caption{Interaction force ${\cal F}$ for two subfamilies of equal
extreme Kerr black holes: ($a$) the case of attracting black holes
($-1/\sqrt{2}<p<0$, $q>0$), ($b$) the case of repelling black
holes ($0<p<1/\sqrt{2}$, $q<0$).}
\end{figure}

\begin{figure}[htb]
\centerline{\epsfysize=50mm\epsffile{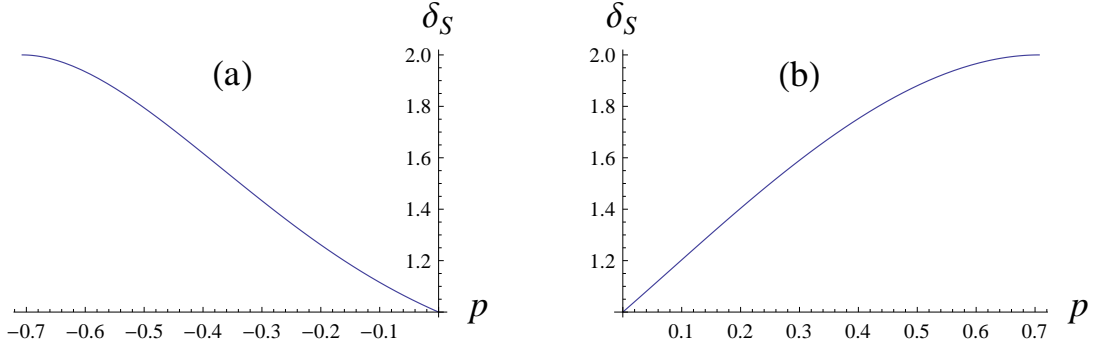}} \caption{Plot of
the angular-momentum--mass ratio $\delta_S(p)$ for two subfamilies
of equal extreme Kerr black holes from section~II: ($a$) the case
of attracting black holes; ($b$) the case of repelling black
holes. For both subfamilies, $\delta_S$ runs the values on the
interval ($1,2$).}
\end{figure}

\begin{figure}[htb]
\centerline{\epsfysize=80mm\epsffile{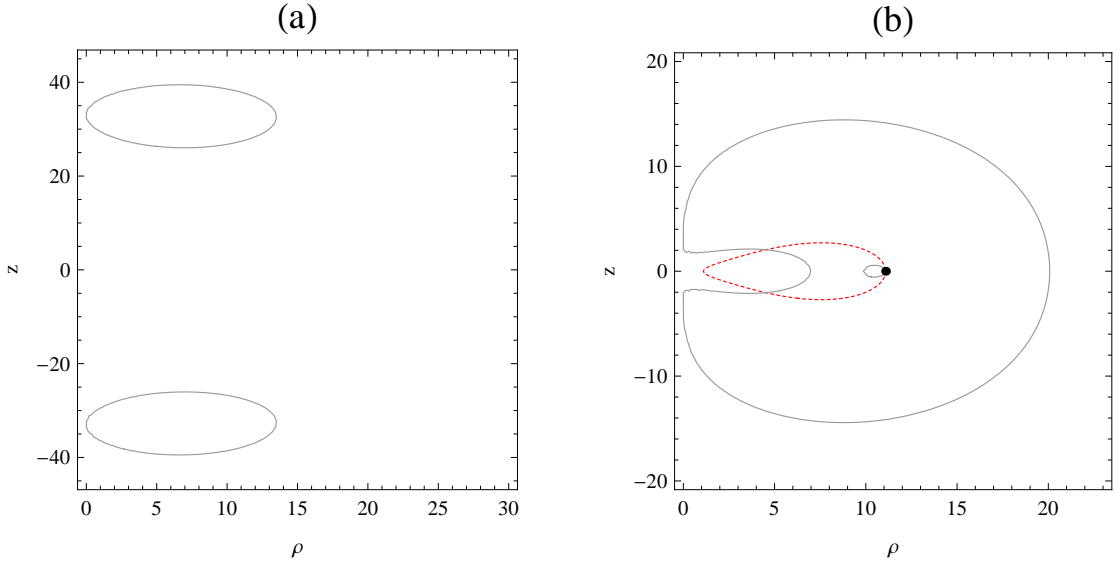}} \caption{The SLSs
for two different binary configurations of extreme Kerr black
holes that have the same masses and angular momenta
($M\simeq13.11$ and $J\simeq240.628$): ($a$) the attracting black
holes located at $\kappa\simeq\pm33.075$ which do not develop a
massless ring singularity or the region of CTCs; ($b$) the
repelling black holes located at $\kappa=\pm2$ which develop a
massless ring singularity (at $z=0$, $\rho\simeq11.081$) and the
region with CTCs (inside the dotted curve), thus indicating that
the common SLS tends to be divided into two separate parts.}
\end{figure}

\begin{figure}[htb]
\centerline{\epsfysize=80mm\epsffile{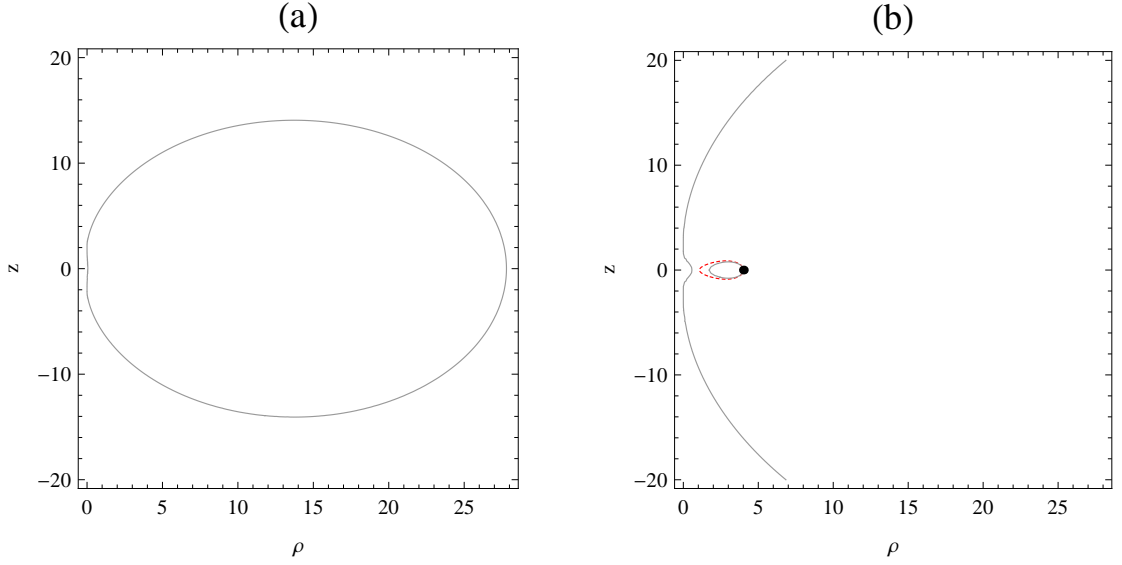}} \caption{The SLSs
for two binary configurations of extreme Kerr black holes located
at $\kappa=\pm2$ and characterized by the same ratio
$\delta_S=1.99$: ($a$) the individual mass and angular momentum of
attracting black holes are $M\simeq13.978$ and $J\simeq388.792$,
respectively, a massless ring singularity off the symmetry axis
and the region of CTCs are absent; ($b$) the repelling black holes
have individual mass $M\simeq30.714$ and angular momentum
$J\simeq1877.263$ and develop a massless ring singularity at
$z=0$, $\rho\simeq4.007$, accompanied by a region of CTCs.}
\end{figure}

\begin{figure}[htb]
\centerline{\epsfysize=80mm\epsffile{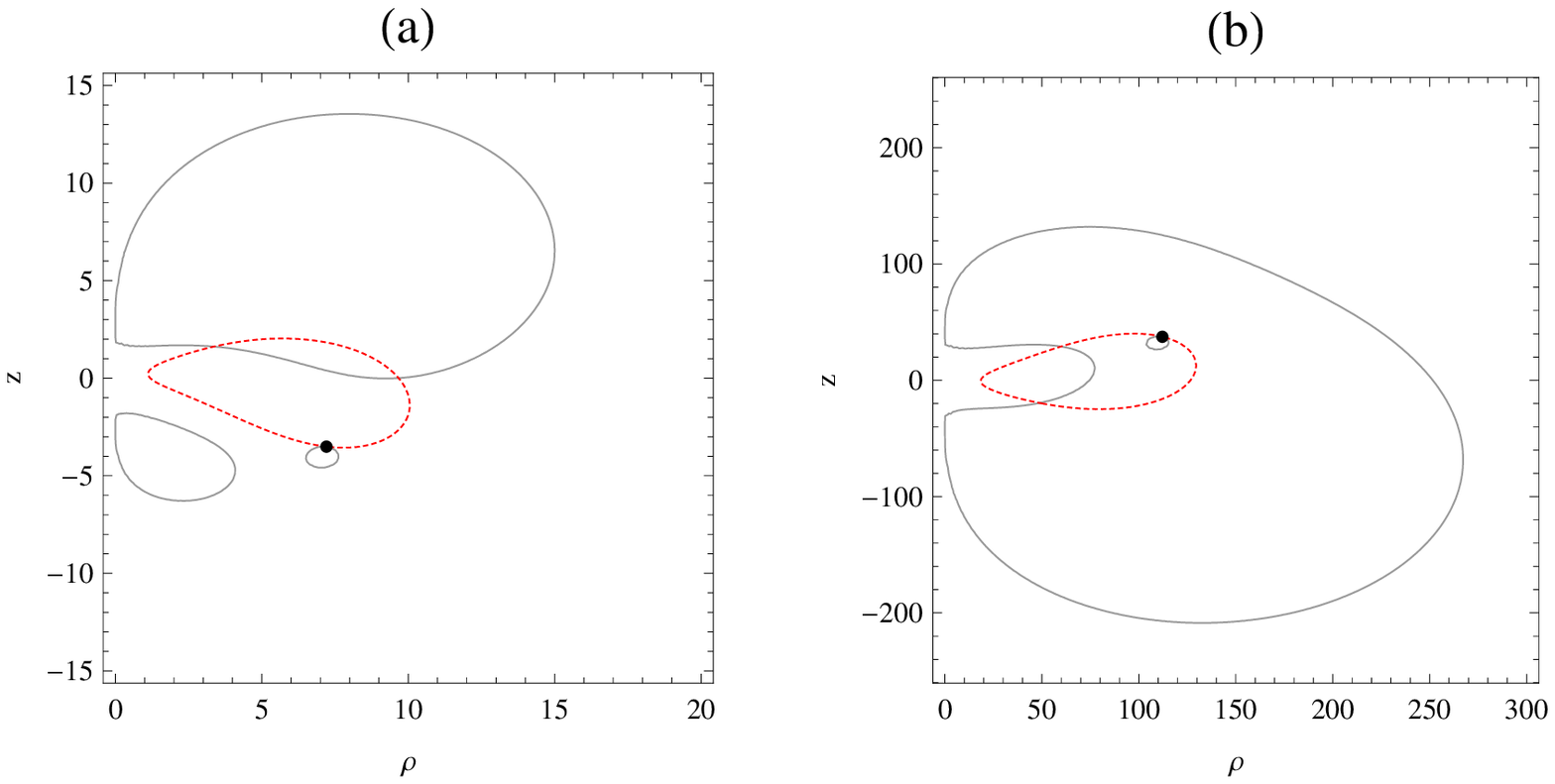}} \caption{The SLSs
for two binary configurations of repelling extreme unequal black
holes defined by the same values of the parameters $\kappa$, $p$
and $q$ as in Fig.~4: ($a$) a configuration from the subfamily
(\ref{pos_2}) with $P=0.5$, $M_T\simeq16.112$ and
$\delta_T\simeq0.945$; the massless ring singularity is located at
$\rho\simeq7.209$, $z\simeq-3.508$, ($b$) a binary system from the
subfamily (\ref{pos_1}) with $P=-0.8$, $M_T\simeq307.381$ and
$\delta_T\simeq0.832$; the massless ring singularity is located at
$\rho\simeq112.126$, $z\simeq37.431$.}
\end{figure}

\begin{figure}[htb]
\centerline{\epsfysize=80mm\epsffile{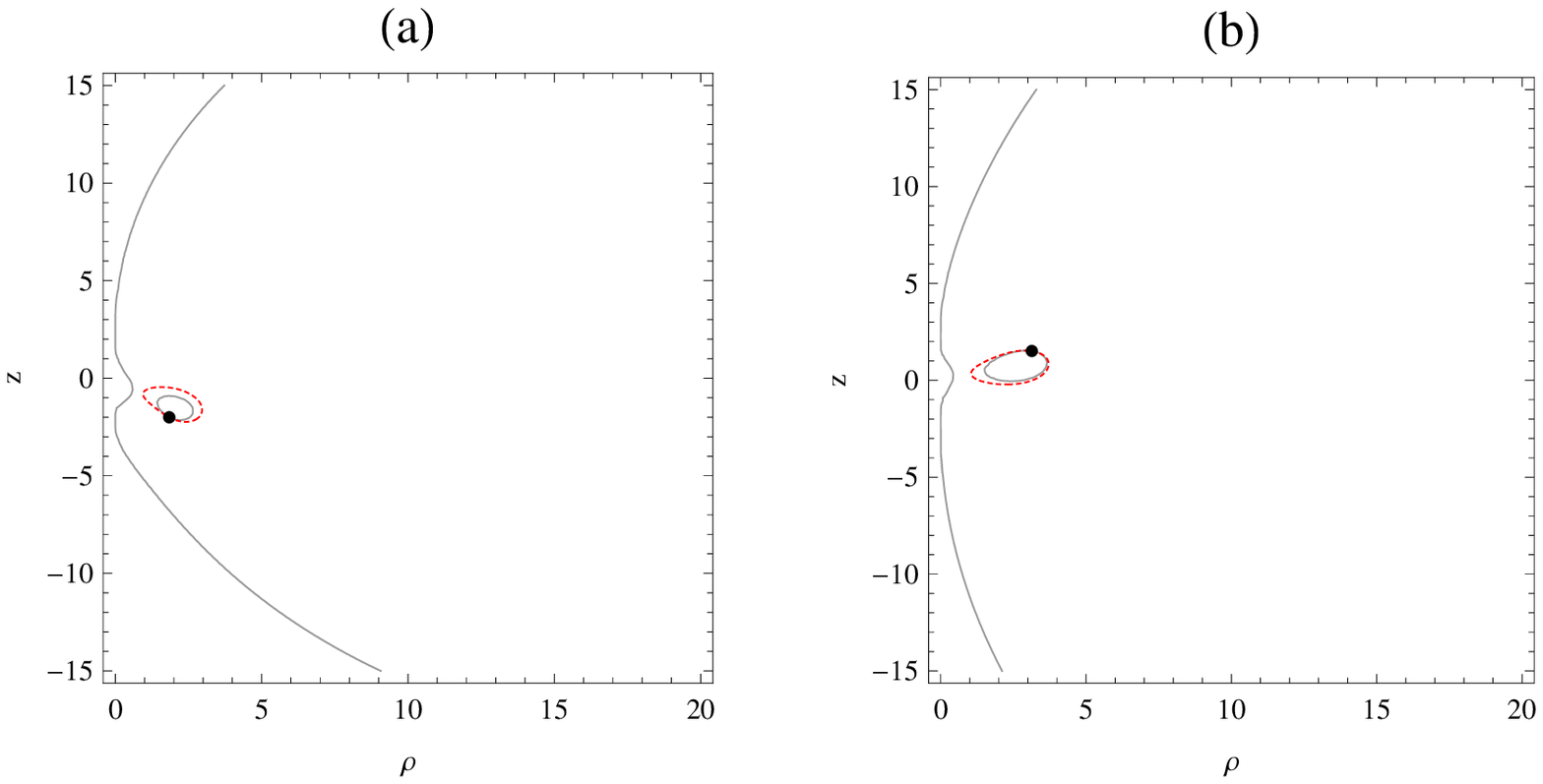}} \caption{The SLSs
for two binary configurations of repelling extreme unequal black
holes defined by the same values of the parameters $\kappa$, $p$
and $q$ as in Fig.~5: ($a$) a configuration from the subfamily
(\ref{pos_2}) with $P=0.2$, $M_T\simeq43.74$ and
$\delta_T\simeq1.00042$; the massless ring singularity is located
at $\rho\simeq1.845$, $z\simeq-2.004$, ($b$) a binary system from
the subfamily (\ref{pos_1}) with $P=-0.8$, $M_T\simeq79.603$ and
$\delta_T\simeq0.998$; the massless ring singularity is located at
$\rho\simeq3.135$, $z\simeq1.512$. }
\end{figure}


\begin{references}

\bibitem{MRu} V. S. Manko and E. Ruiz, ``Black hole-naked singularity'' dualism and the
repulsion of two Kerr black holes due to spin-spin interaction,
\J{\PLB}{791}{26}{2019}.

\bibitem{MRu2} V. S. Manko and E. Ruiz, Extended multi-soliton
solutions of the Einstein field equations,
\J{\CQG}{15}{2007}{1998}.

\bibitem{MRS} V. S. Manko, E. Ruiz and J. D. Sanabria-G\'omez,
Extended multi-soliton solutions of the Einstein field equation:
II. Two comments on the existence of equilibrium states,
\J{\CQG}{17}{3881}{2000}.

\bibitem{KNe} D. Kramer and G. Neugebauer, The
superposition of two Kerr solutions, \J{\PLA}{75}{259}{1980}.

\bibitem{KCh} W. Kinnersley and D. M. Chitre, Symmetries of
the Einstein-Maxwell equations. IV. Transformations which preserve
asymptotic flatness, \J{\JMP}{19}{2037}{1978}.

\bibitem{MRu3} V. S. Manko and E. Ruiz, On a simple representation
of the Kinnersley-Chitre metric, \J{\PTP}{125}{1241}{2011}.

\bibitem{MRSa} V. S. Manko, E. Ruiz, and M. B. Sadovnikova, Stationary
configurations of two extreme black holes obtainable from the
Kinnersley-Chitre solution, \J{\PRD}{84}{064005}{2011}.

\bibitem{Ker} R. P. Kerr, Gravitational field of a spinning mass as
an example of algebraically special metrics,
\J{\PRL}{11}{237}{1963}.

\bibitem{Ern} F. J. Ernst, New formulation of the axially symmetric
gravitational field problem, \J{\PR}{167}{1175}{1968}.

\bibitem{BWe} R. Bach and H. Weyl, Neue L\"osungen der
Einsteinschen Gravitationsgleichungen, \J{Math.
Z.}{13}{134}{1922}.

\bibitem{Isr} W. Israel, Line sources in general relativity,
\J{\PRD}{15}{935}{1977}.

\bibitem{Kom} A. Komar, Covariant conservation laws in
general relativity, \J{\PR}{113}{934}{1959}.

\bibitem{Wei} G. Weinstein, On rotating black holes in
equilibrium in general relativity, \J{\CPAM}{43}{903}{1990}.

\bibitem{Wal} R. Wald, Gravitational spin interaction,
\J{\PRD}{6}{406}{1972}.

\bibitem{MRu4} V. S. Manko and E. Ruiz, Metric for two equal black
holes, \J{\PRD}{96}{104016}{2017}.

\bibitem{Tom} A. Tomimatsu, On gravitational mass and
angular momentum of two black holes in equilibrium,
\J{\PTP}{70}{385}{1983}.

\bibitem{DHo} W. Dietz and C. Hoenselaers, Two mass solutions
of Einstein's vacuum equations: The double Kerr solution,
\J{\APN}{165}{319}{1985}.

\bibitem{SKM} H. Stephani, D. Kramer, M. MacCallum, C.
Hoenselaers, and E. Herlt, Exact Solutions of Einstein's Field
Equations (Cambridge: Cambridge University Press, 2003) p.~307.

\bibitem{TSa} A. Tomimatsu and H. Sato, New exact solution
for the gravitational field of a spinning mass,
\J{\PRL}{29}{1344}{1972}.

\bibitem{GRu} G. W. Gibbons and R. A. Russell-Clark, Note on
the Sato-Tomimatsu solution of Einstein's equations,
\J{\PRL}{30}{398}{1973}.

\bibitem{Hoe} C. Hoenselaers, Remarks on the double-Kerr
solution, \J{\PTP}{72}{761}{1984}.

\bibitem{Man} V. S. Manko, On the physical interpretation
of $\delta=2$ Tomimatsu-Sato solution, \J{\PTP}{127}{1057}{2012}.

\bibitem{GDo} R. J. Gleiser and G. Dotti, Instability
of the negative mass Schwarzschild naked singularity,
\J{\CQG}{23}{5063}{2006}.

\bibitem{CCa} V. Cardoso and M. Cavagli\`a, Stability of naked singularities
and algebraically special modes, \J{\PRD}{74}{024027}{2006}.

\bibitem{DOr} S. Dain and O. E. Ortiz, Numerical evidences for
the angular momentum-mass inequality for multiple axially
symmetric black holes, \J{\PRD}{80}{024045}{2009}.

\bibitem{KZL} S. Komossa, H. Zhou, and H. Lu, A recoiling
supermassive black hole in the quasar SDSS J092712.65+294344.0?
\J{\AJ}{678}{L81}{2008}.

\bibitem{BSK} L. Blecha, D. Sijacki, L. Z. Kelley, {\it et al}, Recoiling black
holes: prospects for detection and implications of spin alignment,
\J{\MNRAS}{456}{961}{2016}.

\end{references}
\end{document}